# Two Sets of Simple Formulae to Estimating Fractal Dimension of Irregular Boundaries

Yanguang Chen

(Department of Geography, College of Urban and Environmental Sciences, Peking University, Beijing 100871, P.R. China. E-mail: chenyg@pku.edu.cn)

**Abstract**: Irregular boundary lines can be characterized by fractal dimension, which provides important information for spatial analysis of complex geographical phenomena such as cities. However, it is difficult to calculate fractal dimension of boundaries systematically when image data is limited. An approximation estimation formulae of boundary dimension based on square is widely applied in urban and ecological studies. However, the boundary dimension is sometimes overestimated. This paper is devoted to developing a series of practicable formulae for boundary dimension estimation using ideas from fractals. A number of regular figures are employed as reference shapes, from which the corresponding geometric measure relations are constructed; from these measure relations, two sets of fractal dimension estimation formulae are derived for describing fractal-like boundaries. Correspondingly, a group of shape indexes can be defined. A finding is that different formulae have different merits and spheres of application, and the second set of boundary dimensions is a function of the shape indexes. Under condition of data shortage, these formulae can be utilized to estimate boundary dimension values rapidly. Moreover, the relationships between boundary dimension and shape indexes are instructive to understand the association and differences between characteristic scales and scaling. The formulae may be useful for the pre-fractal studies in geography, geomorphology, ecology, landscape science, and especially, urban science.

# 1. Introduction

Fractal systems can be characterized by fractal dimension, and the basic and important approach to understanding fractal dimension is the geometric measure relations. Euclidean geometric measure relations come from the principle of dimension consistency. A measure (e.g. length) is not proportional to another measure (e.g. area) unless they share the same spatial dimension (Chen, 2015; Lee, 1989; Mandelbrot, 1983). From the principle of dimensional homogeneity, we can derive Euclidean geometric measure relations, which can be generalized to fractal geometric measure relation (Feder, 1988; Mandelbrot, 1983; Takayasu, 1990). From fractal measure relations, we can derive fractal dimension and allometric scaling exponents (Batty and Longley, 1988; Benguigui *et al*, 2006; Mandelbrot, 1983; Chen and Wang, 2016). An allometric scaling relation can be regarded as a generalized fractal measure relation. Among various geometric measure relations, the common one is the area-perimeter scaling relation, which was used to obtain the boundary dimension of self-similar shapes embedded into a 2-dimensional space (Feder, 1988; Mandelbrot, 1983). In urban studies, the form dimension and boundary dimension can be derived from the fractal measure relations (Batty and Longley, 1994; Frankhauser, 1994). Form dimension of cities include box dimension and radial dimension, which are defined on the basis of the relations of urban area and the linear sizes of box or radius of concentric circles (Batty and Longley, 1994; Frankhauser, 1998). This work focuses on boundary dimension, which can be associated with form dimension in theory.

If we have enough data, we can calculate various fractal parameters. Taking urban research as an example, we can research spatial distribution and structure using box dimension, and research urban growth using radial dimension. If there is not enough information for urban morphology, the boundary dimension of a city can be calculated (Batty and Longley, 1994; Longley and Batty, 1989a; Longley and Batty, 1989b). However, sometimes, we only know the urban area and urban envelope. Urban envelope represents closed urban boundary lines, and urban area represents the region within the boundary curve (Batty and Longley, 1994; Longley *et al*, 1991). In this case, we can only estimate the boundary dimension by means of the data sets of urban area and perimeter length. Referring to a Euclidean shape, we can construct a series of estimation formulae of fractal dimension. The basic reference shapes are standard circle and square. The formula of boundary dimension based on square has been constructed by Olsen *et al* (1993), and the formula was widely applied in urban and ecological studies (Chang, 1996; Chang and Wu, 1998). The formula is simple and easy to understand, and has strong practicability. However, it has two drawbacks. First, this formula is mainly applicable to objects extended in the form of squares. Second, the formula sometimes overestimates the boundary dimension (Chen, 2013; Chen and Wang, 2016). Therefore, we need not only the fractal dimension estimation formulae based on other reference shapes, but also a new fractal dimension estimation formula with reference square. This paper is devoted to deriving two sets of approximation formulae of fractal dimension estimation for the fractal-like boundary dimension of irregular shapes. The reference figures include regular triangle, square, regular hexagon, and standard circle. As a contrast model, the generator of Koch snowflake curve is also employed as one of the reference shapes. From the geometric measure relations based on these reference shapes, a series of formulae are derived to approximately estimate the boundary dimension of various irregular shapes such urban envelopes.

## 2. Fractal measure relations

### 2.1 The first set of formulae

The so-called fractal dimension values based on the approximation formulae are actually fractal indicators, which can be used to replace fractal dimension under the condition of absence adequate data. A basic postulate is that the boundary line of an irregular region is a closed pre-fractal curve. A pre-fractal is a fractal-like object, which is not a real fractal (Addison, 1997; Mitchell, 2009). The "length" of true fractal line is infinite. If the irregular boundary such as urban envelope is a real fractal curve, we cannot derive any simple formula for fractal indicators and shape indexes. Using the ideas from pre-fractals, we can find a number of approximation formulae of boundary dimension from a given reference shape. The reference shapes are some types of regular geometric figures defined in a 2-dimensional Euclidean space, including standard circle, regular triangle, square, regular hexagon, and regular six-pointed star (Figure 1). Triangle can be regarded as the basic shape in Euclidean geometry. All geometric figures, including squares, rectangles, trapezois, circles, ellipses, and irregular shapes, can be reduced to triangles. So the first formula of fractal dimension estimation for irregular boundaries should be derived from a regular triangle. For an equilateral triangle with a side length $r$, the area $A$ and perimeter $P$ can be expressed as follows

$$A = \frac{\sin(\pi/3)}{2} r^2, \qquad (1)$$

$$P = 3r. \qquad (2)$$

Thus the geometric measure relation between area $A$ and perimeter $P$ can be obtained by combining equation (1) with equation (2). Eliminating the side length $r$ yields

$$r = (\frac{2A}{\sin(\pi/3)})^{1/2} = \frac{P}{3}. \qquad (3)$$

Suppose that the three Euclidean sides are replaced by three fractal lines, and the fractal dimension of these lines is $D$. In this case, a regular shape changes to an irregular shape (Figure 2). Thus the Euclidean geometric measure relation, equation (3), should be substituted by a fractal geometric measure relation:

$$(\frac{2A}{\sin(\pi/3)})^{1/2} = (\frac{P}{3})^{1/D}, \qquad (4)$$

where $D$ denotes the fractal dimension of the boundary line. From equation (4), a formula of estimating the fractal dimension of boundary lines can be derived as below:

$$D = \frac{2\ln(P/3)}{\ln(2A/\sin(\pi/3))} = \frac{2\ln(P/3)}{\ln(4A/\sqrt{3})}. \qquad (5)$$

If the shape of a natural system such as a city is similar to a triangle, or a system has three growing directions, the fractal dimension of the system's boundary line can be estimated by equation (5).

The second formula can be constructed on the basis of square. A square is simple and regular, and it is easy to calculate its area $A$ and perimeter $P$ if the side length $r$ is known. The area and perimeter formulae are as follows

$$A = r^2, \qquad (6)$$

$$P = 4r. \qquad (7)$$

Combining equation (7) and equation (6) yields the geometric measure relation between the area $A$ and perimeter $P$ as below:

$$r = A^{1/2} = \frac{P}{4}. \qquad (8)$$

If the sides of the square are replaced by the fractal lines with fractal dimension $D$, the geometric measure relation, equation (8), will be replaced by

$$A^{1/2} = (\frac{P}{4})^{1/D}. \tag{9}$$

From equation (9), a fractal dimension estimation formula can be derived as follows

$$D = \frac{2\ln(P/4)}{\ln(A)}. \tag{10}$$

This formula is familiar to many scholars who like geographical and ecological fractals because it was once derived by Olsen et al (1993) in another way. If the shape of a natural system is similar to a square, or a system has four growing directions, the boundary dimension of the system shape can be estimated by equation (10).

The regular hexagons can be best closed to each other in a geographical region. Therefore, the hexagonal networks were applied to the well-known central place theory (Christaller, 1966). The area $A$ and perimeter $P$ of a regular hexagon with a side length $r$ can be calculated by the following formulae:

$$A = 3\sin(\frac{\pi}{3})r^2 = \frac{3\sqrt{3}}{2}r^2, \tag{11}$$

$$P = 6r. \tag{12}$$

From the equations (11) and (12), we can derive a geometric measure relation such as

$$r = (\frac{2A}{3\sqrt{3}})^{1/2} = (\frac{A}{3\sin(\pi/3)})^{1/2} = \frac{P}{6}. \tag{13}$$

Substituting the Euclidean sides of the hexagon with fractal boundary lines, we can turn equation (13) into a fractal measure relation as follows

$$r = (\frac{2A}{3\sqrt{3}})^{1/2} = (\frac{A}{3\sin(\pi/3)})^{1/2} = (\frac{P}{6})^{1/D}. \tag{14}$$

From equation (14), a fractal dimension estimation formula can be obtained as below:

$$D = \frac{2\ln(P/6)}{\ln(2A/(3\sqrt{3}))} = \frac{2\ln(P/6)}{\ln(A/(3\sin(\pi/3)))}. \tag{15}$$

If the shape of a natural system is similar to a hexagon, or a system has six growing directions, the boundary dimension of the system can be estimated by equation (15).

The standard circle is treated a simple and perfect shape in Euclidean geometry. Many shape indexes of geography are based on this kind of circle (Chen, 2011; Haggett et al, 1977; Lin, 1998; Taylor, 1977). The area $A$ and perimeter $P$ of a circle with a radius $r$ can be given by

$$A = \pi r^2, \tag{16}$$

$$P = 2\pi r. \tag{17}$$

Integrating equation (16) into equation (17) yields the geometric measure relation between the circular area $A$ and circumference $P$ as follows

$$r = (\frac{A}{\pi})^{1/2} = \frac{P}{2\pi}. \tag{18}$$

Replacing the Euclidean perimeter with a fractal curve results in a fractal measure relation as below:

$$(\frac{A}{\pi})^{1/2} = (\frac{P}{2\pi})^{1/D}. \tag{19}$$

Thus the fractal dimension of the boundary can be calculated by the following formula

$$D = \frac{2\ln(P/(2\pi))}{\ln(A/\pi)}. \tag{20}$$

If the shape of a natural system is similar to a circle, or if a system is of isotropic growth, the boundary dimension of the system can be estimated by equation (20).

All the above-given formulae are based on Euclidean figures. For comparison, it is advisable to construct a calculation formula based on fractal generators. Koch snowflake curve is one of classical models for fractal lines. We can design the formula using the generator of Koch snowflake curve, a regular six-pointed star. For fractal generator of Koch snowflake curve with a side length $r$, the area $A$ and perimeter $P$ are as follows

$$A = 6\sin(\pi/3)r^2, \tag{21}$$

$$P = 12r. \tag{22}$$

Thus the geometric measure relation between the area $A$ and perimeter $P$ can be derived as

$$r = (\frac{A}{6\sin(\pi/3)})^{1/2} = \frac{P}{12}. \tag{23}$$

The generator of Koch curve is not a fractal line, but the second step is a pre-fractal figure. Substituting the straight line segments with fractal lines yields a fractal measure relation as below:

$$(\frac{A}{6\sin(\pi/3)})^{1/2} = (\frac{P}{12})^{1/D}, \tag{24}$$

in which $D$ refers to the fractal index of irregular curve. The formula of fractal dimension estimation based on the Koch snowflake generator can be derived from equation (24) as follows

$$D = \frac{2\ln(P/12)}{\ln(A/(6\sin(\pi/3)))} = \frac{2\ln(P/12)}{\ln(A/(3\sqrt{3}))}. \quad (25)$$

If the shape of a natural system is similar to a Koch snowflake, or a system has six protruding growth directions, the boundary dimension of the system can be estimated by equation (25).

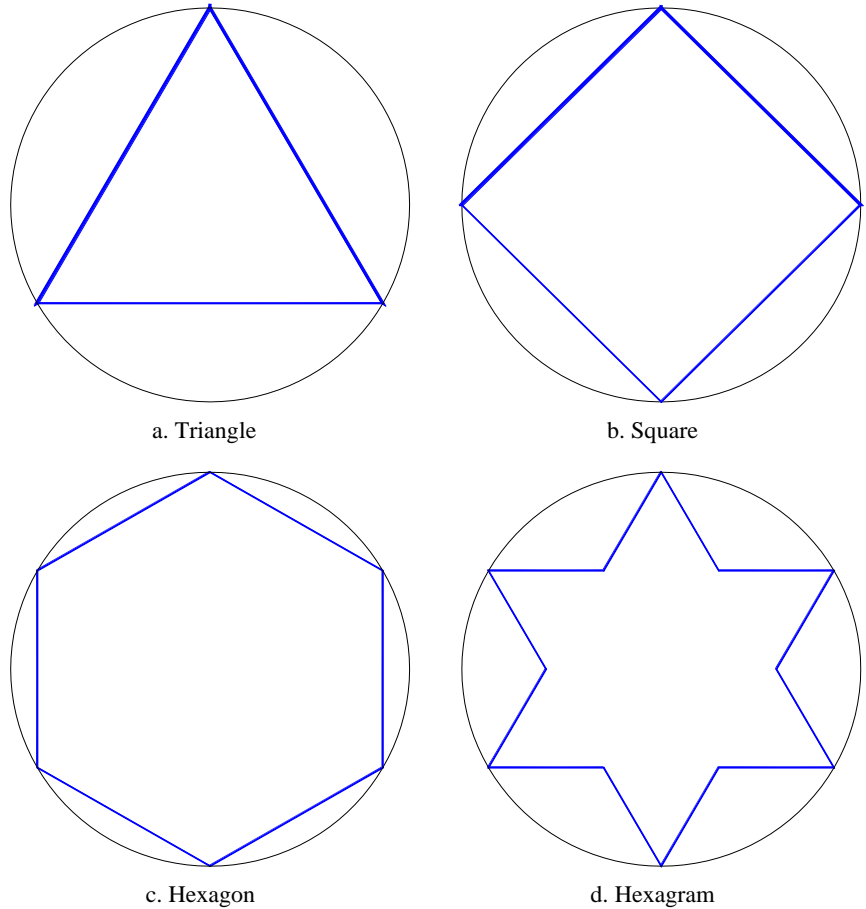

a. Triangle   b. Square

c. Hexagon   d. Hexagram

**Figure 1 Four typical reference shapes for derivation of approximation formulae of fractal dimension of irregular boundaries**

Note: The circumcircles of these shapes represent the standard circles, from which we can derive the approximation formulae of boundary dimension and the criterion values for shape indexes.

## 2.2 The second set of formulae

The above-shown formulae are suitable for the very irregular boundary lines, otherwise the fractal dimension may be overestimated. In order to estimate the boundary dimension of general fractal-like line, we should improve the formulae. Based on regular triangles, equation (4) can be replaced by the following fractal measure relation

$$\left(\frac{2A}{\sin(\pi/3)}\right)^{1/2} = \frac{P^{1/D}}{3}, \tag{26}$$

which can be expressed as

$$\left(\frac{18A}{\sin(\pi/3)}\right)^{1/2} = (12\sqrt{3}A)^{1/2} = P^{1/D}. \tag{27}$$

From equations (26) and (27), a triangle-based fractal dimension formula can be derived as below:

$$D = \frac{2\ln(P)}{\ln(24\sin(\pi/3)A)} = \frac{2\ln(P)}{\ln(12\sqrt{3}A)}. \tag{28}$$

Compared with equation (5), equation (28) can give more realistic dimension values of fractal lines. Based on squares, equation (9) can be substituted by

$$A^{1/2} = \frac{P^{1/D}}{4}, \tag{29}$$

which can be rewritten as

$$(16A)^{1/2} = P^{1/D}. \tag{30}$$

From equation (30), a square-based fractal dimension formula can be derived as follows

$$D = \frac{2\ln(P)}{\ln(16A)}. \tag{31}$$

Compared with equation (10), equation (31) can yield more realistic values of boundary dimension. Based on regular hexagons, equation (14) can be replaced with

$$r = \left(\frac{2A}{3\sqrt{3}}\right)^{1/2} = \left(\frac{A}{3\sin(\pi/3)}\right)^{1/2} = \frac{P^{1/D}}{6}. \tag{32}$$

which is equivalent to

$$r = \left(\frac{24A}{\sqrt{3}}\right)^{1/2} = \left(\frac{12A}{\sin(\pi/3)}\right)^{1/2} = P^{1/D}. \tag{33}$$

From equations (32) and (33), a hexagon-based fractal dimension formula can be derived as follows

$$D = \frac{2\ln(P)}{\ln(12A/\sin(\pi/3))} = \frac{2\ln(P)}{\ln(8\sqrt{3}A)}. \tag{34}$$

Compared with equation (15), equation (34) can produce more realistic values of boundary dimension. Based on the standard circle, equation (19) can be replaced by

$$(\frac{A}{\pi})^{1/2} = \frac{P^{1/D}}{2\pi}, \tag{35}$$

which can be converted into

$$(4\pi A)^{1/2} = P^{1/D}. \tag{36}$$

From equation (36), a circle-based fractal dimension formula can be derived as below

$$D = \frac{2\ln(P)}{\ln(4\pi A)}. \tag{37}$$

Compared with equation (20), equation (37) can give more realistic fractal dimension values of boundary lines. Based on the regular six-pointed star, equation (24) can be substituted with

$$(\frac{A}{6\sin(\pi/3)})^{1/2} = \frac{P^{1/D}}{12}, \tag{38}$$

which can be converted into

$$(\frac{24A}{\sin(\pi/3)})^{1/2} = P^{1/D}. \tag{39}$$

From equations (38) and (39), a Koch-snowflake-based fractal dimension formula can be derived as below

$$D = \frac{2\ln(P)}{\ln(24A/\sin(\pi/3))} = \frac{2\ln(P)}{\ln(16\sqrt{3}A)}. \tag{40}$$

Compared with equation (25), equation (40) can give more realistic values of fractal indexes of boundary lines.

The two sets of fractal dimension estimation formulae represent two sets of fractal boundary indexes. For the convenience of readers, the two sets of formulae are tabulated as follows (Table 1). Applying these formulae to a simple irregular shape (Figure 2), we can obtain two sets of fractal indexes values for boundary lines (Table 2). Based on the first set of formulae, the fractal dimension estimation results are marked as boundary dimension $D_{(1)}$; and based on the second set of formulae, the fractal dimension estimation results are marked as boundary dimension $D_{(2)}$. The values of $D_{(2)}$ is less than those of $D_{(1)}$.

**Table 1 The summary of the main simple formulae for estimating fractal dimension of irregular closed boundary curves**

| Initiator | Reference shape | The formula for higher fractal dimension | The formula for lower fractal dimension | Shape index |
|---|---|---|---|---|
| Euclidean shape | Regular triangle | $D = \dfrac{2\ln(P/3)}{\ln(4A/\sqrt{3})}$ | $D = \dfrac{2\ln(P)}{\ln(12\sqrt{3}A)}$ | $s = \dfrac{12\sqrt{3}A}{P^2}$ |
| | Square | $D = \dfrac{2\ln(P/4)}{\ln(A)}$ | $D = \dfrac{2\ln(P)}{\ln(16A)}$ | $s = \dfrac{16A}{P^2}$ |
| | Regular hexagon | $D = \dfrac{2\ln(P/6)}{\ln(2A/(3\sqrt{3}))}$ | $D = \dfrac{2\ln(P)}{\ln(8\sqrt{3}A)}$ | $s = \dfrac{8\sqrt{3}A}{P^2}$ |
| | Standard circle | $D = \dfrac{2\ln(P/(2\pi))}{\ln(A/\pi)}$ | $D = \dfrac{2\ln(P)}{\ln(4\pi A)}$ | $s = \dfrac{4\pi A}{P^2}$ |
| Fractal shape | Regular six-pointed star | $D = \dfrac{2\ln(P/12)}{\ln(A/(3\sqrt{3}))}$ | $D = \dfrac{2\ln(P)}{\ln(16\sqrt{3}A)}$ | $s = \dfrac{16\sqrt{3}A}{P^2}$ |

**Note**: The first formula based on square was proposed by Olsen *et al* (1993). As a reference, the corresponding shape indexes are listed in the right column.

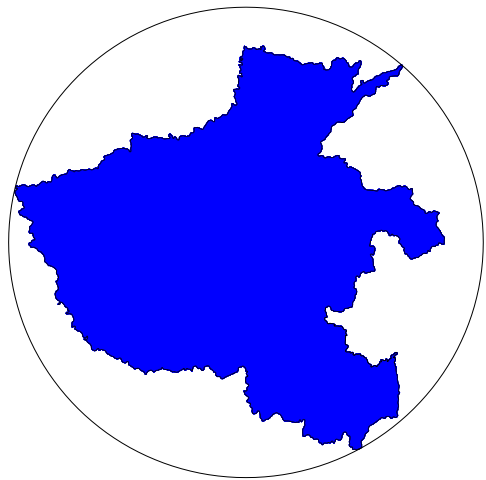

**Figure 2 A simple irregular shape with area *A*=1515.5368 unit and perimeter *P*=295.1157 unit**

Note: The boundary curve of this irregular region bears the property of fractal line. Using different approximation formulae, we can obtain different fractal dimension values.

## 2.3 Relations of boundary dimension to shape indexes

The area-perimeter measure relation is in essence a problem of scaling in complex systems. The formulae of boundary dimension represent various definitions of the scaling exponents of different shapes. The traditional thinking of mathematical modeling and quantitative analysis is based on characteristic scales (Chen, 2008; Hao, 1986; Liu and Liu, 1993; Takayasu, 1990; Wang and Li, 1996). Therefore, a number of shape indexes have been derived from the area-perimeter relations to describe natural morphology such as urban patterns (Chen, 2011; Haggett *et al*, 1977; Lin, 1998; Taylor, 1977). Two typical shape indexes are the circularity ratios, which are defined on the basis of area and perimeter (Haggett *et al*, 1977). However, complex systems such as cities have no significant characteristic scales in many aspects (Chen, 2008). In this case, the scaling concept is employed to substitute the notion of typical scales such as characteristic lengths. Correspondingly, we utilize scaling exponents instead of shape indexes to characterize the form features of irregular patterns. The scaling exponents are based on the ideas from fractal geometry, while the shape indexes are based on the notion of Euclidean geometry. Despite the difference between the scaling exponents and shape indexes, there are inherent association of boundary dimension with shape indexes. The relations between the circularity ratios and the reciprocal of the boundary dimension has been proved to be the exponential function (Chen, 2011). Based on area and perimeter, a series of shape indexes similar to the circularity ratios can be derived from the above-shown geometric measure relations, which are listed in Table 1. Applying these formulae to Figure 2 yields a series of values for shape indexes. The relations between the new shapes indexes and the second set of boundary dimension can be demonstrated to satisfy the following exponential function

$$s = \frac{1}{P^2}\exp(\frac{2\ln(P)}{D}) = a\exp(\frac{b}{D}), \quad (41)$$

where the parameters are $a=1/P^2$ and $b=2\ln(P)$. For example, for Figure 2, the relationship between the second type of boundary dimension and the shape index is as follows

$$s = 0.00001148\exp(\frac{11.3747}{D}), \quad (42)$$

which can be verified by the observational data (Table 2). It is easy to testify that $a=1/295.1157^{\wedge}2 = 0.00001148$ and $b = 2*\ln(295.1157) =11.3747$.

If the boundary line is a true fractal line, then the perimeter will be infinite. In this case, equation (41) will be invalid. This suggests that the fractal dimensions and shape indexes in Table 1 will be invalid for true fractal boundary lines. In the real world, the boundary lines are fractal-like lines rather than true fractal lines. A real fractal bears no scaling limitation, and the geometric measure relation can be reflected by a straight line of infinite length on a log-log plot. In contrast, a pre-fractal possesses fractal nature only within certain scaling range. Therefore, the perimeters of pre-fractal closed curves bear certain values, and thus the formulae derived above are valid. Moreover, equation (41) lend further support to the inference that the relationships between the reciprocal of boundary dimension and the circularity ratios meet an exponential function. However, the relations between the shapes indexes and the first set of boundary dimension cannot be described with the exponential function exactly. Despite this, the inherent correlation between characteristic scales and scaling can be reflected by the relationships between boundary dimension and shape indexes.

Table 2 The boundary dimension values and the corresponding shape index values

| Reference shape | Boundary dimension $D_{(1)}$ | Boundary dimension $D_{(2)}$ | Shape index $s$ |
|---|---|---|---|
| Regular triangle | 1.1246 | 1.0982 | 0.3617 |
| Square | 1.1746 | 1.1266 | 0.2784 |
| Regular hexagon | 1.2234 | 1.1429 | 0.2411 |
| Standard circle | 1.2460 | 1.1543 | 0.2187 |
| Regular six-pointed star | 1.1285 | 1.0685 | 0.4822 |

**Note**: The boundary dimension values are estimated for Figure 2, for which, the area is $A$=1515.5368 unit and perimeter is $P$=295.1157 unit.

## 3. Empirical analysis

### 3.1 Fractal dimension estimation results

To show the effects of the two sets of fractal dimension estimation formulae, we can apply them to the cities in Beijing, Tianjin, and Hebei region, China (for short, Jing-Jin-Ji region). There are 13 main cities in the study area. As a preparation, it is necessary to extract urban boundary lines using a proper method. In urban geography, the boundary curve of a city is termed *urban envelope*, and the region within the urban envelope is termed *urban area* (Batty and Longley, 1994; Longley *et al*, 1991). There are at least four scientific approaches to identifying and delineating urban envelopes for these cities (Chen *et al*, 2019), including the city clustering algorithm (CCA) (Rozenfeld *et al*, 2008), the automatic identification method of urban settlement boundaries (Chaudhry and Mackaness, 2008), the fractal-based method (Tannier *et al*, 2011), and the approach to derive 'natural cities' by clustering street nodes/blocks (Jiang and Jia, 2011). In this paper, CCA is employed to delineate urban boundary lines on interpreted remote sensing images in different years. The urban boundary determined by this method corresponds to an urban agglomeration, which approximately corresponds to an urbanized area. The urban envelopes give urban perimeters, and the corresponding urban areas can be counted (Table 3). These datasets can be used to verify the geometric measure relation between urban area and perimeter and evaluate the fractal boundary indexes (Chen and Wang, 2016; Chen *et al*, 2019).

Table 3 The measures of area and perimeter of major cities in Beijing-Tianjin-Hebei Region in 2000, 2005, and 2010

| City | 2000 | | 2005 | | 2010 | |
|---|---|---|---|---|---|---|
| | Perimeter | Area | Perimeter | Area | Perimeter | Area |
| Baoding | 648.9145 | 165.6927 | 614.5161 | 177.6204 | 618.4115 | 181.7597 |
| Beijing | 1851.1617 | 1633.7361 | 2638.7369 | 2372.6209 | 3256.7105 | 2890.8456 |
| Cangzhou | 359.7652 | 96.5368 | 411.0804 | 111.0267 | 387.1366 | 107.9078 |
| Chengde | 383.2282 | 83.3696 | 386.8770 | 82.2602 | 385.7290 | 85.9915 |
| Handan | 497.2757 | 165.9174 | 587.4799 | 176.9860 | 587.4799 | 176.9860 |
| Hengshui | 286.5358 | 86.7690 | 217.3211 | 81.6262 | 275.6421 | 109.7244 |
| Langfang | 288.8174 | 96.1140 | 292.5072 | 99.8148 | 285.4678 | 101.9814 |
| Qinhuangdao | 429.0707 | 121.1679 | 451.5457 | 159.4474 | 403.8776 | 173.9307 |
| Shijiazhuang | 671.2512 | 329.6986 | 796.1045 | 389.8407 | 852.5650 | 446.2061 |
| Tangshan | 720.8655 | 214.2018 | 744.8371 | 226.1832 | 771.5667 | 264.4586 |
| Tianjin | 1867.5855 | 850.6298 | 2428.3672 | 1595.2827 | 2480.1692 | 2029.3596 |
| Xingtai | 402.1714 | 125.8745 | 391.3236 | 128.2210 | 386.5214 | 132.7910 |
| Zhuangjiakou | 181.2180 | 53.7583 | 186.2037 | 55.4021 | 249.4361 | 90.7300 |
| Average | 660.6047 | 309.4974 | 780.5308 | 435.1025 | 841.5933 | 522.5133 |

**Note**: The results based on the data in 2000, 2005, and 2010 are partially shown below. All the results can be found in the supplementary files.

It is easy to calculate the boundary fractal dimension of the 13 cities in Beijing, Tianjin and Hebei region using each formula. As indicated above, the fractal dimension estimation values based on the first set of formulae are marked as boundary dimension $D_{(1)}$; and the fractal dimension estimation values based on the second set of formulae are marked as boundary dimension $D_{(2)}$. Based on the first set of formulae, the estimated values of the boundary fractal dimension are sometimes very high or very low, or even greater than 2 or less than 0. In contrast, if we utilize the second set of formulae to estimate the boundary dimension, the results are relatively reasonable. All the values range from 1 to 2. In other words, no value is greater than 2 or less than 0. The results of 2000, 2005, and 2010 are listed below (Table 4, Table 5, Table 6). The shape indexes are listed separately for reference (Table 7). The boundary fractal dimension based on the first set of formulae has weak correlation with the shape index. However, there is negative correlation between the fractal dimension values based on the second set of formulae and the shape indexes (Supplementary File 1).

Table 4 Two sets of fractal dimension estimation for boundary lines of major cities in Beijing-Tianjin-Hebei Region in 2000

| City | Boundary dimension $D_{(1)}$ | | | | | Boundary dimension $D_{(2)}$ | | | | |
|---|---|---|---|---|---|---|---|---|---|---|
| | Triangle | Square | Hexagon | Circle | Hexagram | Triangle | Square | Hexagon | Circle | Hexagram |
| Baoding | 1.8082 | 1.9917 | 2.2542 | 2.3389 | 2.3051 | 1.5901 | 1.6429 | 1.6734 | 1.6948 | 1.5359 |
| Beijing | 1.5603 | 1.6590 | 1.7790 | 1.8183 | 1.7524 | 1.4423 | 1.4794 | 1.5006 | 1.5154 | 1.4036 |
| Cangzhou | 1.7706 | 1.9690 | 2.2647 | 2.3634 | 2.3275 | 1.5480 | 1.6031 | 1.6351 | 1.6577 | 1.4915 |
| Chengde | 1.8440 | 2.0629 | 2.3969 | 2.5077 | 2.4960 | 1.5953 | 1.6533 | 1.6871 | 1.7108 | 1.5361 |
| Handan | 1.7183 | 1.8871 | 2.1254 | 2.2039 | 2.1505 | 1.5245 | 1.5751 | 1.6044 | 1.6249 | 1.4725 |
| Hengshui | 1.7204 | 1.9141 | 2.2039 | 2.3022 | 2.2541 | 1.5093 | 1.5638 | 1.5956 | 1.6179 | 1.4535 |
| Langfang | 1.6908 | 1.8747 | 2.1458 | 2.2380 | 2.1805 | 1.4910 | 1.5442 | 1.5751 | 1.5968 | 1.4367 |
| Qinhuangdao | 1.7618 | 1.9492 | 2.2225 | 2.3128 | 2.2715 | 1.5480 | 1.6015 | 1.6326 | 1.6543 | 1.4932 |
| Shijiazhuang | 1.6309 | 1.7671 | 1.9480 | 2.0077 | 1.9393 | 1.4739 | 1.5189 | 1.5448 | 1.5630 | 1.4274 |
| Tangshan | 1.7672 | 1.9356 | 2.1707 | 2.2465 | 2.2025 | 1.5666 | 1.6169 | 1.6460 | 1.6664 | 1.5147 |
| Tianjin | 1.6969 | 1.8222 | 1.9825 | 2.0333 | 1.9802 | 1.5403 | 1.5827 | 1.6070 | 1.6239 | 1.4963 |
| Xingtai | 1.7271 | 1.9071 | 2.1673 | 2.2539 | 2.2037 | 1.5241 | 1.5765 | 1.6069 | 1.6282 | 1.4703 |
| Zhuangjiakou | 1.7012 | 1.9141 | 2.2497 | 2.3677 | 2.3237 | 1.4817 | 1.5390 | 1.5725 | 1.5961 | 1.4233 |
| Average | 1.7229 | 1.8964 | 2.1470 | 2.2303 | 2.1836 | 1.5258 | 1.5767 | 1.6062 | 1.6269 | 1.4735 |

Table 5 Two sets of fractal dimension estimation for boundary lines of major cities in Beijing-Tianjin-Hebei Region in 2005

| City | Boundary dimension $D_{(1)}$ | | | | | Boundary dimension $D_{(2)}$ | | | | |
|---|---|---|---|---|---|---|---|---|---|---|
| | Triangle | Square | Hexagon | Circle | Hexagram | Triangle | Square | Hexagon | Circle | Hexagram |
| Baoding | 1.7692 | 1.9440 | 2.1913 | 2.2716 | 2.2289 | 1.5634 | 1.6148 | 1.6446 | 1.6654 | 1.5105 |
| Beijing | 1.5750 | 1.6706 | 1.7856 | 1.8229 | 1.7614 | 1.4581 | 1.4943 | 1.5149 | 1.5293 | 1.4203 |
| Cangzhou | 1.7741 | 1.9672 | 2.2514 | 2.3455 | 2.3083 | 1.5544 | 1.6088 | 1.6403 | 1.6625 | 1.4988 |
| Chengde | 1.8523 | 2.0734 | 2.4117 | 2.5238 | 2.5150 | 1.6008 | 1.6591 | 1.6930 | 1.7168 | 1.5412 |
| Handan | 1.7553 | 1.9279 | 2.1719 | 2.2513 | 2.2057 | 1.5531 | 1.6043 | 1.6338 | 1.6545 | 1.5006 |
| Hengshui | 1.6349 | 1.8151 | 2.0825 | 2.1756 | 2.1033 | 1.4473 | 1.5001 | 1.5308 | 1.5524 | 1.3934 |
| Langfang | 1.6837 | 1.8648 | 2.1306 | 2.2209 | 2.1612 | 1.4870 | 1.5397 | 1.5704 | 1.5919 | 1.4330 |

| Qinhuangdao | 1.6972 | 1.8638 | 2.0991 | 2.1771 | 2.1192 | 1.5082 | 1.5585 | 1.5876 | 1.6080 | 1.4565 |
| Shijiazhuang | 1.6408 | 1.7746 | 1.9509 | 2.0086 | 1.9430 | 1.4844 | 1.5288 | 1.5544 | 1.5723 | 1.4384 |
| Tangshan | 1.7623 | 1.9283 | 2.1589 | 2.2332 | 2.1881 | 1.5642 | 1.6142 | 1.6430 | 1.6632 | 1.5127 |
| Tianjin | 1.6309 | 1.7380 | 1.8702 | 1.9124 | 1.8544 | 1.4977 | 1.5363 | 1.5584 | 1.5738 | 1.4575 |
| Xingtai | 1.7119 | 1.8885 | 2.1430 | 2.2279 | 2.1739 | 1.5136 | 1.5655 | 1.5956 | 1.6167 | 1.4603 |
| Zhuangjiakou | 1.7018 | 1.9133 | 2.2453 | 2.3617 | 2.3171 | 1.4830 | 1.5402 | 1.5736 | 1.5970 | 1.4249 |
| Average | 1.7069 | 1.8746 | 2.1148 | 2.1948 | 2.1446 | 1.5166 | 1.5665 | 1.5954 | 1.6157 | 1.4652 |

**Table 6 Two sets of fractal dimension estimation for boundary lines of major cities in Beijing-Tianjin-Hebei Region in 2010**

| City | Boundary dimension $D_{(1)}$ | | | | | Boundary dimension $D_{(2)}$ | | | | |
|---|---|---|---|---|---|---|---|---|---|---|
| | Triangle | Square | Hexagon | Circle | Hexagram | Triangle | Square | Hexagon | Circle | Hexagram |
| Baoding | 1.7645 | 1.9378 | 2.1824 | 2.2619 | 2.2180 | 1.5606 | 1.6118 | 1.6414 | 1.6621 | 1.5079 |
| Beijing | 1.5875 | 1.6820 | 1.7953 | 1.8318 | 1.7729 | 1.4702 | 1.5060 | 1.5264 | 1.5406 | 1.4327 |
| Cangzhou | 1.7615 | 1.9535 | 2.2364 | 2.3305 | 2.2904 | 1.5446 | 1.5988 | 1.6303 | 1.6524 | 1.4891 |
| Chengde | 1.8357 | 2.0515 | 2.3794 | 2.4881 | 2.4731 | 1.5905 | 1.6481 | 1.6815 | 1.7051 | 1.5316 |
| Handan | 1.7553 | 1.9279 | 2.1719 | 2.2513 | 2.2057 | 1.5531 | 1.6043 | 1.6338 | 1.6545 | 1.5006 |
| Hengshui | 1.6334 | 1.8020 | 2.0450 | 2.1283 | 2.0552 | 1.4534 | 1.5043 | 1.5339 | 1.5546 | 1.4013 |
| Langfang | 1.6681 | 1.8456 | 2.1048 | 2.1932 | 2.1292 | 1.4765 | 1.5287 | 1.5590 | 1.5803 | 1.4230 |
| Qinhuangdao | 1.6354 | 1.7892 | 2.0026 | 2.0744 | 2.0031 | 1.4650 | 1.5133 | 1.5412 | 1.5608 | 1.4153 |
| Shijiazhuang | 1.6287 | 1.7578 | 1.9263 | 1.9816 | 1.9149 | 1.4775 | 1.5210 | 1.5461 | 1.5636 | 1.4323 |
| Tangshan | 1.7303 | 1.8869 | 2.1011 | 2.1704 | 2.1190 | 1.5440 | 1.5924 | 1.6203 | 1.6398 | 1.4941 |
| Tianjin | 1.5895 | 1.6886 | 1.8089 | 1.8478 | 1.7867 | 1.4679 | 1.5048 | 1.5259 | 1.5406 | 1.4292 |
| Xingtai | 1.6971 | 1.8700 | 2.1177 | 2.2005 | 2.1428 | 1.5038 | 1.5551 | 1.5849 | 1.6058 | 1.4511 |
| Zhuangjiakou | 1.6541 | 1.8336 | 2.0981 | 2.1892 | 2.1219 | 1.4636 | 1.5162 | 1.5467 | 1.5682 | 1.4098 |
| Average | 1.6878 | 1.8482 | 2.0746 | 2.1499 | 2.0948 | 1.5054 | 1.5542 | 1.5824 | 1.6022 | 1.4552 |

It is necessary to discuss the exceptional values in the fractal dimension estimation results. In theory, the boundary dimension defined in a 2-dimensional embedding space is supposed to come between 0 and 2 (Batty and Longley, 1994; Chen, 2018; Chen, 2019; Mandelbrot, 1983; Longley and Batty, 1989a; Longley and Batty, 1989b). The reasonable values vary from 1 to 1.5. However, the following causes often result in over estimation of boundary dimension. First, the boundary dimension calculation is based on the geometric measure relation deriving from regular real fractals in the mathematical world. A real fractal has no scaling range, or, in other words, the scaling range of a real fractal is infinite. Applying the fractal measure relations proceeding from regular real fractals to the random pre-fractals gives rise to significant bias in many cases (Chen, 2013). Second, if the image resolution is high enough, the length of the boundary line may be very long, but the area within the boundary curve is certain. This phenomenon can be illustrated by a regular fractal termed Koch lake (see Appendix). The Koch lake can be treated as models of lakes, islands, urban region, and so on. Third, compared with the second set of approximate formulae, the first set of approximate formulae enlarge the ratio of the perimeter logarithm to the area logarithm of a shape relatively. For example, for the formulae based on square, in equation (10), the circumference is reduced to a quarter of the original length; while in equation (31), the area is enlarged to 16 times the original one. As a result, the boundary dimension value of a shape based on equation (10) is significantly greater than the value based on equation (31). Generally speaking, the estimated value of a boundary dimension is not less than 1. However, if a figure is near a Euclidean shape, the fractal dimension estimation result may be an outlier and less than 1 because that the formulae is designed for pre-fractals rather than for Euclidean shapes.

**Table 7 The shape index values of major cities in Beijing-Tianjin-Hebei Region in 2000, 2005, and 2010**

| City | Shape index in 2000 | | | | | Shape index in 2005 | | | | |
|---|---|---|---|---|---|---|---|---|---|---|
| | Triangle | Square | Hexagon | Circle | Hexagram | Triangle | Square | Hexagon | Circle | Hexagram |
| Baoding | 0.0082 | 0.0063 | 0.0055 | 0.0049 | 0.0109 | 0.0098 | 0.0075 | 0.0065 | 0.0059 | 0.0130 |
| Beijing | 0.0099 | 0.0076 | 0.0066 | 0.0060 | 0.0132 | 0.0071 | 0.0055 | 0.0047 | 0.0043 | 0.0094 |
| Cangzhou | 0.0155 | 0.0119 | 0.0103 | 0.0094 | 0.0207 | 0.0137 | 0.0105 | 0.0091 | 0.0083 | 0.0182 |
| Chengde | 0.0118 | 0.0091 | 0.0079 | 0.0071 | 0.0157 | 0.0114 | 0.0088 | 0.0076 | 0.0069 | 0.0152 |
| Handan | 0.0139 | 0.0107 | 0.0093 | 0.0084 | 0.0186 | 0.0107 | 0.0082 | 0.0071 | 0.0064 | 0.0142 |
| Hengshui | 0.0220 | 0.0169 | 0.0146 | 0.0133 | 0.0293 | 0.0359 | 0.0277 | 0.0239 | 0.0217 | 0.0479 |
| Langfang | 0.0239 | 0.0184 | 0.0160 | 0.0145 | 0.0319 | 0.0242 | 0.0187 | 0.0162 | 0.0147 | 0.0323 |
| Qinhuangdao | 0.0137 | 0.0105 | 0.0091 | 0.0083 | 0.0182 | 0.0163 | 0.0125 | 0.0108 | 0.0098 | 0.0217 |
| Shijiazhuang | 0.0152 | 0.0117 | 0.0101 | 0.0092 | 0.0203 | 0.0128 | 0.0098 | 0.0085 | 0.0077 | 0.0170 |
| Tangshan | 0.0086 | 0.0066 | 0.0057 | 0.0052 | 0.0114 | 0.0085 | 0.0065 | 0.0056 | 0.0051 | 0.0113 |
| Tianjin | 0.0051 | 0.0039 | 0.0034 | 0.0031 | 0.0068 | 0.0056 | 0.0043 | 0.0037 | 0.0034 | 0.0075 |
| Xingtai | 0.0162 | 0.0125 | 0.0108 | 0.0098 | 0.0216 | 0.0174 | 0.0134 | 0.0116 | 0.0105 | 0.0232 |
| Zhuangjiakou | 0.0340 | 0.0262 | 0.0227 | 0.0206 | 0.0454 | 0.0332 | 0.0256 | 0.0221 | 0.0201 | 0.0443 |
| Average | 0.0152 | 0.0117 | 0.0102 | 0.0092 | 0.0203 | 0.0159 | 0.0122 | 0.0106 | 0.0096 | 0.0212 |

Continued Table 7

| City | Shape index in 2010 | | | | |
|---|---|---|---|---|---|
| | Triangle | Square | Hexagon | Circle | Hexagram |
| Baoding | 0.0099 | 0.0076 | 0.0066 | 0.0060 | 0.0132 |
| Beijing | 0.0057 | 0.0044 | 0.0038 | 0.0034 | 0.0076 |
| Cangzhou | 0.0150 | 0.0115 | 0.0100 | 0.0090 | 0.0200 |
| Chengde | 0.0120 | 0.0092 | 0.0080 | 0.0073 | 0.0160 |
| Handan | 0.0107 | 0.0082 | 0.0071 | 0.0064 | 0.0142 |
| Hengshui | 0.0300 | 0.0231 | 0.0200 | 0.0181 | 0.0400 |
| Langfang | 0.0260 | 0.0200 | 0.0173 | 0.0157 | 0.0347 |
| Qinhuangdao | 0.0222 | 0.0171 | 0.0148 | 0.0134 | 0.0295 |
| Shijiazhuang | 0.0128 | 0.0098 | 0.0085 | 0.0077 | 0.0170 |
| Tangshan | 0.0092 | 0.0071 | 0.0062 | 0.0056 | 0.0123 |
| Tianjin | 0.0069 | 0.0053 | 0.0046 | 0.0041 | 0.0091 |
| Xingtai | 0.0185 | 0.0142 | 0.0123 | 0.0112 | 0.0246 |
| Zhuangjiakou | 0.0303 | 0.0233 | 0.0202 | 0.0183 | 0.0404 |
| Average | 0.0161 | 0.0124 | 0.0107 | 0.0097 | 0.0214 |

## 3.2 Fractal dimension adjustment and transformation

Fractal lines falls into two types: one is boundary lines, and the other is what is called space-filling curves. In theory, the well-known boundary line is Koch snowflake curve, and the well-known space-filling curve is Peano curve (Mandelbrot, 1983). In reality, fractal boundary lines include coast lines, urban boundaries, lake boundaries, and national boundaries, and space-filling curves include rivers, traffic networks, and hierarchical boundaries in central place systems. Generally speaking, the average values of fractal dimension of boundary lines come between 1 and 1.5, while the fractal dimension values of space-filling curves vary from 1.5 to 2. If a fractal dimension of urban boundary is over estimated, it can be adjusted by a simple formula. What is more, the adjusted boundary dimension can be converted into the fractal dimension of urban form. The adjusted boundary dimension is as follows (Chen, 2013)

$$D_b = \frac{1+D_l}{2}, \qquad (43)$$

where $D_b$ denotes the adjusted boundary dimension, and $D_l$ refers to the originally estimated fractal dimension (Note: The subscript for $D_l$ is the Latin letter l, not the Arabic numeral 1). Using this formula, we can adjust the boundary dimension values (Table 8). The boundary dimension and the form dimension satisfy a hyperbolic relation as blow (Chen, 2013):

$$D_f = 1 + \frac{1}{D_l}, \qquad (44)$$

where $D_f$ refers to the form dimension of a city. Using this formula, we can transform the adjusted boundary dimension values into the fractal dimension of urban form (Table 9).

**Table 8 The adjusted boundary dimension of major cities in Beijing-Tianjin-Hebei Region in 2000, 2005, and 2010 (based on the second set of formulae)**

| City | Adjusted boundary dimension in 2000 | | | | | Adjusted boundary dimension in 2005 | | | | |
|---|---|---|---|---|---|---|---|---|---|---|
| | Triangle | Square | Hexagon | Circle | Hexagram | Triangle | Square | Hexagon | Circle | Hexagram |
| Baoding | 1.2951 | 1.3215 | 1.3367 | 1.3474 | 1.2679 | 1.2817 | 1.3074 | 1.3223 | 1.3327 | 1.2553 |
| Beijing | 1.2211 | 1.2397 | 1.2503 | 1.2577 | 1.2018 | 1.2290 | 1.2471 | 1.2575 | 1.2647 | 1.2101 |
| Cangzhou | 1.2740 | 1.3016 | 1.3176 | 1.3288 | 1.2458 | 1.2772 | 1.3044 | 1.3202 | 1.3312 | 1.2494 |
| Chengde | 1.2977 | 1.3267 | 1.3435 | 1.3554 | 1.2680 | 1.3004 | 1.3295 | 1.3465 | 1.3584 | 1.2706 |
| Handan | 1.2623 | 1.2876 | 1.3022 | 1.3124 | 1.2363 | 1.2766 | 1.3021 | 1.3169 | 1.3273 | 1.2503 |
| Hengshui | 1.2546 | 1.2819 | 1.2978 | 1.3089 | 1.2268 | 1.2237 | 1.2500 | 1.2654 | 1.2762 | 1.1967 |
| Langfang | 1.2455 | 1.2721 | 1.2875 | 1.2984 | 1.2183 | 1.2435 | 1.2699 | 1.2852 | 1.2959 | 1.2165 |
| Qinhuangdao | 1.2740 | 1.3008 | 1.3163 | 1.3272 | 1.2466 | 1.2541 | 1.2793 | 1.2938 | 1.3040 | 1.2283 |
| Shijiazhuang | 1.2370 | 1.2595 | 1.2724 | 1.2815 | 1.2137 | 1.2422 | 1.2644 | 1.2772 | 1.2862 | 1.2192 |
| Tangshan | 1.2833 | 1.3085 | 1.3230 | 1.3332 | 1.2573 | 1.2821 | 1.3071 | 1.3215 | 1.3316 | 1.2564 |
| Tianjin | 1.2702 | 1.2913 | 1.3035 | 1.3119 | 1.2482 | 1.2489 | 1.2682 | 1.2792 | 1.2869 | 1.2287 |
| Xingtai | 1.2620 | 1.2882 | 1.3034 | 1.3141 | 1.2352 | 1.2568 | 1.2828 | 1.2978 | 1.3084 | 1.2302 |
| Zhuangjiakou | 1.2408 | 1.2695 | 1.2863 | 1.2980 | 1.2117 | 1.2415 | 1.2701 | 1.2868 | 1.2985 | 1.2124 |
| Average | 1.2629 | 1.2884 | 1.3031 | 1.3135 | 1.2367 | 1.2583 | 1.2833 | 1.2977 | 1.3078 | 1.2326 |

Continued Table 8

| City | Adjusted boundary dimension in 2010 | | | | |
|---|---|---|---|---|---|
| | Triangle | Square | Hexagon | Circle | Hexagram |
| Baoding | 1.2803 | 1.3059 | 1.3207 | 1.3311 | 1.2540 |
| Beijing | 1.2351 | 1.2530 | 1.2632 | 1.2703 | 1.2164 |
| Cangzhou | 1.2723 | 1.2994 | 1.3152 | 1.3262 | 1.2446 |
| Chengde | 1.2952 | 1.3240 | 1.3408 | 1.3525 | 1.2658 |
| Handan | 1.2766 | 1.3021 | 1.3169 | 1.3273 | 1.2503 |
| Hengshui | 1.2267 | 1.2522 | 1.2669 | 1.2773 | 1.2006 |
| Langfang | 1.2382 | 1.2643 | 1.2795 | 1.2901 | 1.2115 |
| Qinhuangdao | 1.2325 | 1.2566 | 1.2706 | 1.2804 | 1.2076 |
| Shijiazhuang | 1.2387 | 1.2605 | 1.2730 | 1.2818 | 1.2162 |
| Tangshan | 1.2720 | 1.2962 | 1.3101 | 1.3199 | 1.2470 |
| Tianjin | 1.2339 | 1.2524 | 1.2630 | 1.2703 | 1.2146 |
| Xingtai | 1.2519 | 1.2776 | 1.2924 | 1.3029 | 1.2255 |
| Zhuangjiakou | 1.2318 | 1.2581 | 1.2734 | 1.2841 | 1.2049 |
| Average | 1.2527 | 1.2771 | 1.2912 | 1.3011 | 1.2276 |

**Table 8 The estimated form dimension of major cities in Beijing-Tianjin-Hebei Region in 2000**

**and 2010 (based on the adjusted boundary dimension)**

| City | Form dimension in 2000 | | | | | Form dimension in 2005 | | | | |
|---|---|---|---|---|---|---|---|---|---|---|
| | Triangle | Square | Hexagon | Circle | Hexagram | Triangle | Square | Hexagon | Circle | Hexagram |
| Baoding | 1.7722 | 1.7567 | 1.7481 | 1.7422 | 1.7887 | 1.7802 | 1.7649 | 1.7563 | 1.7503 | 1.7967 |
| Beijing | 1.8189 | 1.8067 | 1.7998 | 1.7951 | 1.8321 | 1.8136 | 1.8018 | 1.7952 | 1.7907 | 1.8263 |
| Cangzhou | 1.7849 | 1.7683 | 1.7590 | 1.7525 | 1.8027 | 1.7829 | 1.7666 | 1.7575 | 1.7512 | 1.8004 |
| Chengde | 1.7706 | 1.7538 | 1.7443 | 1.7378 | 1.7886 | 1.7690 | 1.7521 | 1.7427 | 1.7362 | 1.7870 |
| Handan | 1.7922 | 1.7767 | 1.7679 | 1.7619 | 1.8089 | 1.7834 | 1.7680 | 1.7594 | 1.7534 | 1.7998 |
| Hengshui | 1.7970 | 1.7801 | 1.7705 | 1.7640 | 1.8152 | 1.8172 | 1.8000 | 1.7903 | 1.7836 | 1.8356 |
| Langfang | 1.8029 | 1.7861 | 1.7767 | 1.7702 | 1.8208 | 1.8042 | 1.7875 | 1.7781 | 1.7716 | 1.8220 |
| Qinhuangdao | 1.7849 | 1.7688 | 1.7597 | 1.7535 | 1.8022 | 1.7974 | 1.7817 | 1.7729 | 1.7669 | 1.8142 |
| Shijiazhuang | 1.8084 | 1.7940 | 1.7859 | 1.7803 | 1.8239 | 1.8050 | 1.7909 | 1.7830 | 1.7775 | 1.8202 |
| Tangshan | 1.7793 | 1.7643 | 1.7559 | 1.7501 | 1.7953 | 1.7800 | 1.7651 | 1.7567 | 1.7510 | 1.7959 |
| Tianjin | 1.7873 | 1.7744 | 1.7672 | 1.7622 | 1.8012 | 1.8007 | 1.7885 | 1.7817 | 1.7771 | 1.8139 |
| Xingtai | 1.7924 | 1.7762 | 1.7672 | 1.7610 | 1.8096 | 1.7957 | 1.7796 | 1.7705 | 1.7643 | 1.8129 |
| Zhuangjiakou | 1.8059 | 1.7877 | 1.7775 | 1.7704 | 1.8253 | 1.8055 | 1.7873 | 1.7771 | 1.7701 | 1.8248 |
| Average | 1.7921 | 1.7764 | 1.7677 | 1.7616 | 1.8088 | 1.7950 | 1.7795 | 1.7709 | 1.7649 | 1.8115 |

Continued Table 9

| City | Form dimension in 2010 | | | | |
|---|---|---|---|---|---|
| | Triangle | Square | Hexagon | Circle | Hexagram |
| Baoding | 1.7811 | 1.7658 | 1.7572 | 1.7513 | 1.7975 |
| Beijing | 1.8097 | 1.7981 | 1.7916 | 1.7872 | 1.8221 |
| Cangzhou | 1.7860 | 1.7696 | 1.7604 | 1.7540 | 1.8035 |
| Chengde | 1.7721 | 1.7553 | 1.7458 | 1.7394 | 1.7900 |
| Handan | 1.7834 | 1.7680 | 1.7594 | 1.7534 | 1.7998 |
| Hengshui | 1.8152 | 1.7986 | 1.7893 | 1.7829 | 1.8329 |
| Langfang | 1.8076 | 1.7909 | 1.7816 | 1.7751 | 1.8254 |
| Qinhuangdao | 1.8114 | 1.7958 | 1.7870 | 1.7810 | 1.8281 |
| Shijiazhuang | 1.8073 | 1.7933 | 1.7855 | 1.7802 | 1.8223 |
| Tangshan | 1.7862 | 1.7715 | 1.7633 | 1.7576 | 1.8019 |
| Tianjin | 1.8104 | 1.7985 | 1.7918 | 1.7872 | 1.8233 |
| Xingtai | 1.7988 | 1.7827 | 1.7737 | 1.7675 | 1.8160 |
| Zhuangjiakou | 1.8118 | 1.7949 | 1.7853 | 1.7788 | 1.8299 |
| Average | 1.7985 | 1.7833 | 1.7748 | 1.7689 | 1.8148 |

The approximate fractals dimensions are in fact fractal indexes. An index can condense many data into a number, describing the characteristics of a system and simplifying the analytical process. Given the area and perimeter of an irregular shape such as an urban envelope, we can calculate its fractal indexes using these sets of formulae. All these approximate formulae can be applied to estimation of boundary dimension of the cities in the Yangtze River Delta, China. The datasets in 1985, 1996, and 2005 have been published (Chen and Wang, 2016). The results show that the fractal index values of city boundaries in the Yangtze River Delta are significantly lower than those in Beijing, Tianjin, and Hebei region. The reason may be that the resolution of remote sensing images of Beijing, Tianjin and Hebei cities is higher than that of Yangtze River Delta cities (Supplementary File 2).

## 4. Discussion

The above-shown results are based on power-law relations, and a power law represents a geometric measure relation and reflects a proportional relationship. A power function has two parameters: one is the *proportionality coefficient*, and the other is the *power exponent*. In the framework of Euclidean geometry, the power exponent is always a known constant and bears little useful information. Thus we can construct various shape indexes based on proportionality coefficients. A proportional constant is always a dimensionless parameter reflecting a ratio of one measure to another measure. On the contrary, in the framework of fractal geometry, the proportionality coefficient bears little information, but the power exponent is unknown parameter and possesses spatial information. A simple system has characteristic scale and can be described with the mathematical method based on Euclidean geometry, while a complex system has no characteristic scale and cannot be effectively described by conventional mathematical methods. In this case, it is necessary to replace the characteristic scale concept with scaling idea. The power exponent is known as scaling exponent. Fractal geometry is a powerful tool for scaling analysis of complex systems, and the fractal dimension is an important scaling exponent. Based on the notion of fractals, various fractal indexes can be defined to characterize fractal-like phenomena.

A set of formulae for estimating the boundary dimension of irregular shapes have been derived from the geometric measure relations between the area and perimeter of certain reference figures. The reference shapes include regular triangle, square, regular hexagon, circle, and the generator of Koch snowflake curve. From difference reference shape, we can obtain different formulae; from the same reference shape, we can obtain at least two different formula based on different conditions (Table 1). Different formulae have different spheres of application. In practice, we should select the proper formula according to the shape and irregularity of studied objects. As a matter of fact, a number of methods can be used to calculate the boundary dimension. The common methods include divider method (Mandelbrot, 1967; Richardson, 1961), box counting method (Song et al, 2012; Wang et al, 2005), and so on. Longley and Batty (1989a; 1989b) developed four methods to measure the fractal dimension of fractal lines. Given enough image data, it is not problematic to calculate the boundary dimension (Batty and Longley, 1988; Chen and Wang, 2016; Cheng, 1995; Imre, 2006; Imre and Bogaert, 2004). The formulae proposed in this paper are suitable for fractal dimension estimation of irregular boundaries under the condition of data shortage. Concretely speaking, we need to use these formulas in three cases. First, limitation of data. The amount of data is small, and the existing data do not support the calculation of fractal dimension by the least squares calculation. Second, approximation of results. An approximate estimation of fractal dimension can meet the needs of special research. Third, comparability of datasets. Image data of different years or places have the same quality. For example, we don't have any data except the numbers of boundary lengths of a city's and the area within the boundary lines in different years. In this instance, we can estimate the boundary dimension of the city and analyze its growing process and pattern. The results of fractal dimension estimation are not real fractal dimension, but a kind of characteristic indexes to describe the fractal-like boundaries.

A fractal measure relation can be treated as an allometric scaling relation. These scaling relations are widely applied to urban research. In fact, the geometric measure relation between urban area and urban perimeter bears analogy with the allometric scaling relation between urban population and urban area, which can be expressed as

$$S^{1/D_p} \propto A^{1/2}, \tag{45}$$

where $S$ denotes the population size of a city. Thus the allometric scaling relation between urban population size $S$ and urban boundary length $P$ can be derived as follows

$$S^{1/D_p} \propto P^{1/D}, \tag{46}$$

which can be re-expressed as

$$S \propto P^{D_p/D} = \mu P^{\alpha}, \tag{47}$$

where $\mu = SP^{-\alpha}$ represents the proportionality coefficient, and $\alpha = D_p/D$ is the scaling exponent. This suggests that urban population size is in a proportion to *α* power of the urban perimeter.

Further, the geometric measure relation can be generalized to the traffic network of a city. Suppose that the urban area is *A* and the total length of traffic lines is *L*. According to the principle of dimension consistency, we have

$$L^{1/D_w} \propto A^{1/2}, \quad (48)$$

where $D_w$ denotes the fractal dimension of traffic networks. For comparability, the proportionality coefficient is assumed to be 1, then equation (48) can be transformed into

$$2\ln(L) = D_w \ln(A). \quad (49)$$

From equation (49) it follows

$$D_w = \frac{2\ln(L)}{\ln(A)}. \quad (50)$$

This is the formula of fractal dimension of traffic network in an urbanized area. This also implies that the effect of measures' dimension on the result can be eliminated by taking logarithms of measures when constructing a fractal index.

Fractal measures are significant in the research on complex landscape, which bear no characteristic scale and cannot be characterized by the common indexes in theory. Cities represent complex human landscape. The shape indexes and boundary dimension are basic measures in urban studies, and both characteristic scales and scaling are important concepts in urban geography. Comparably speaking, scaling concept is more important. Cities are complex spatial systems, and many aspects of urban systems have no characteristic scales. Scaling in cities has attracted more and more attention of scholars (Arcaute *et al*, 2015; Batty, 2008; Bettencourt, 2013; Bettencourt *et al*, 2007; Chen, 2008; Louf and Barthelemy, 2014a; Louf and Barthelemy, 2014b; Ortman *et al*, 2014; Pumain, 2006). Fractal geometry is one of powerful tools in scaling analysis and has been applied to urban studies, which resulted in a number of interesting achievements. The series of approximation formulae of boundary dimension provide simple approaches to scaling analysis of cities. The shortcomings of this study lies in two aspects. First, the formulae of boundary dimensions and the corresponding shape indexes have not yet be derived from regular pentagon. Compared with the regular triangle, square, regular hexagon, and standard circle, the regular pentagon is more complex because it can be associated with fractals. Second, the systematic positive study has not been made. The empirical evidences shown in this paper is only for methodology rather than for urban studies. What is more, the formulae are only suitable for random pre-fractals rather than real fractals (Appendix). Due to the limitation of space of a paper, the pending questions will be answered in future studies.

## 5. Conclusions

Boundary line represents a perspective of spatial patterns and landscape analysis of complex systems. In this paper, various possible formulae for estimating boundary dimension are systematically studied and compared. The aim of this work is to provide two sets of practical formulae for approximate estimation of boundary dimension of fractal-like phenomena. The main conclusions can be reached as follows. First, *the approximate formulae of boundary dimension can be derived from geometric measure relation in light of scaling thinking*. Traditional mathematical modeling and quantitative analysis are based on characteristic scales, and a number of shape indexes are derived from these relations to describe various shapes; complex systems bear no characteristic scales, so a number of scaling exponents are derived to characterize various patterns. Second, *the approximate estimation formulae of boundary dimension are not unique, but diverse*. On the one hand, different fractal boundary dimension can be defined based on different reference shapes; on the other, under different conditions, different formulae can be derived from the same geometric measure relation. Therefore, in practice, proper approximate formula should be selected according to the shape and irregularity of natural morphology. Third, *the approximate boundary dimension are essentially scaling exponents for describing complex curves*. These formulas are used to estimate fractal parameters of boundary lines only in case of data shortage. The estimation results using the approximation formulae are actually fractal indexes instead of the calculated values of real fractal dimension. These formulae are suitable for pre-fractal curves rather than real fractal lines. If we have enough image data, we should use normal methods to calculate the fractal dimension with higher confidence level. Sometimes, even if the data is sufficient, these formulas can be used to estimate the fractal parameters quickly when the accuracy requirement is not so high.

**Acknowledgements**

This research was sponsored by the National Natural Science Foundations of China (Grant No.

41590843 & 41671167). The supports are gratefully acknowledged. The author would like to thank Ms Yuqing Long of Peking University for providing the essential data on China's cities.

## Supplementary files

File 1. Datasets for the fractal indexes based on urban area-perimeter relations of Beijing-Tianjin-Hebei Region

File 2. Datasets for the fractal indexes based on urban area-perimeter relations of Yangtze River Delta

## Appendix--Fractal indexes of Koch lake

The approximate formulae developed in this paper is suitable for fractal-like curves rather than real fractal lines. Let's apply the fractal index formulae to a regular fractal, i.e., Koch lake, which comprises of three Koch curves (Figure A). Suppose that the initiator of each Koch curve is a line segment of unit length, namely, $L_1=1$. Thus the initiator of Koch lake is an equilateral triangle with a perimeter $P_1=3L=3$. Correspondingly, the area of the triangle is $A_1= L_1^2\sin(\pi/3)/2= \cos(\pi/3)\sin(\pi/3) = 3^{1/2}/4$. According to the knowledge of geometry and trigonometry, the perimeter of Koch lake can be expressed as

$$P_m = 3(\frac{1}{3^{m-1}})^{1-D} = 3(\frac{1}{3^{m-1}})^{1-\ln(4)/\ln(3)} = 3(\frac{1}{3^{m-1}})*(3^{m-1})^{\log_3 4} = 3(\frac{4}{3})^{m-1}, \qquad (A1)$$

where $m=1,2,3,\ldots$ denotes steps of fractal generation. Clearly, if $m\to\infty$, then $P_m\to\infty$. In contrast, the area of Koch lake is limited. In light of the knowledge of geometry and trigonometry, the area of Koch lake is

$$A_m = A_1(1+\frac{1}{3}\sum_{i=1}^{m}((\frac{1}{3})^2 4)^{i-1}) = \sin(\frac{\pi}{3})\cos(\frac{\pi}{3})(1+\frac{1}{3}\sum_{i=1}^{m}(\frac{4}{9})^{i-1}), \qquad (A2)$$

where $m=1,2,\ldots$ refers to steps of fractal development. Under the limit condition, the area of Koch lake approximates a constant, that is,

$$A_m = \frac{\sqrt{3}}{4}(1+\frac{1}{3}\lim_{m\to\infty}\sum_{i=1}^{m}(\frac{4}{9})^{i-1}) \to \frac{\sqrt{3}}{4}(1+\frac{1}{3(1-4/9)}) = \frac{2\sqrt{3}}{5}. \qquad (A3)$$

This is one of the characteristics of fractals: infinite filling in a finite space. In this case, regardless of the formula in Table 1, the fractal dimension is a variable dependent on $m$ instead of a constant.

For example, based on square, the fractal indexes and shape index are as follows

$$D_m = \frac{2(m-2)\ln(\frac{4}{3})}{\ln(\frac{2\sqrt{3}}{5})}, \quad D_m = \frac{2\ln(3(\frac{4}{3})^{m-1})}{\ln(\frac{32\sqrt{3}}{5})}, \quad s_m = \frac{32\sqrt{3}}{5}/(9(\frac{4}{3})^{2(m-1)}).$$

This suggests that both the approximate fractal dimension formulae and shape index are meaningless for real fractal curves.

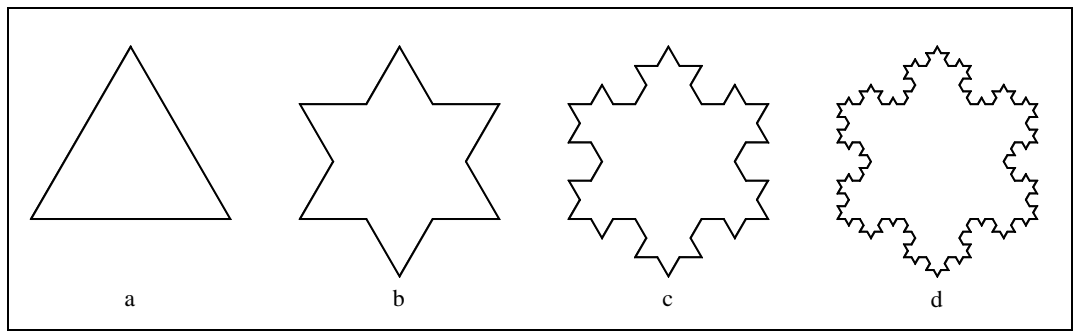

**Figure A The first four steps of the Koch lake model**
**Note**: Koch's lake is sometimes termed Koch island in literature.